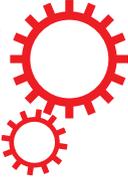

# SCIENTIFIC REPORTS

OPEN

# A micro-scale simulation of red blood cell passage through symmetric and asymmetric bifurcated vessels

Tong Wang[1], Uwitije Rongin[1] & Zhongwen Xing[2]

Blood exhibits a heterogeneous nature of hematocrit, velocity, and effective viscosity in microcapillaries. Microvascular bifurcations have a significant influence on the distribution of the blood cells and blood flow behavior. This paper presents a simulation study performed on the two-dimensional motions and deformation of multiple red blood cells in microvessels with diverging and converging bifurcations. Fluid dynamics and membrane mechanics were incorporated. Effects of cell shape, hematocrit, and deformability of the cell membrane on rheological behavior of the red blood cells and the hemodynamics have been investigated. It was shown that the blood entering the daughter branch with a higher flow rate tended to receive disproportionally more cells. The results also demonstrate that red blood cells in microvessels experienced lateral migration in the parent channel and blunted velocity profiles in both straight section and daughter branches, and this effect was influenced by the shape and the initial position of the cells, the hematocrit, and the membrane deformability. In addition, a cell free region around the tip of the confluence was observed. The simulation results are qualitatively consistent with existing experimental findings. This study may provide fundamental knowledge for a better understanding of hemodynamic behavior of micro-scale blood flow.

Microvascular system is a network of microvessels connected by irregular short bifurcated segments. The local blood flow behavior at the vicinity of the apexes of the diverging and converging bifurcation significantly affects the hemodynamics in microcirculation, such as the heterogeneous distribution of red blood cells throughout the microvasculature, the effective viscosity, and blood flow velocity[1]. The direct influence of this nonuniform distribution is on the microvascular oxygen transportation. Therefore, detailed investigation and deeper understanding of rheological property of red blood cells in micro-bifurcations are extremely important.

Blood flow in microcirculation can be affected by many factors. One of the most crucial factors is hematocrit, which is the volumetric percentage of red blood cells in whole blood, and it ranges from 37% to 52% for adults. Due to the Fahraeus-Lindqvist effect, the tube hematocrit in smaller branches can be as low as 10–20%, much lower than the hematocrit in the main vessels. The variation in hematocrit may cause fluctuation in blood effective viscosity which is a primary measure of blood flow resistance and a crucial factor of friction against the vessel walls. Thus, an increase in the blood viscosity will lead to decreasing oxygen transported to tissues and organs and will cause problems with blood circulation.

Another important factor that affects the microscopic blood flow is the deformability of the red blood cell. Under normal conditions, red blood cells deform into various shapes to compensate the flow resistance. The cell membrane can be significantly altered under some disease conditions. For example, the malaria infected red blood cells could be much more rigid in comparison with healthy ones[2,3]. The ability that red blood cells possess to deform themselves influences the aggregation of the cells, the cell free layer, as well as the velocity of the cells to transit the vessels. The mechanical property of the red blood cell membrane has been modeled and the effects on blood flow have been extensively studied[4–6].

Due to the complexity of the microvascular system, a detailed *in vivo* study of blood flow in the microcirculation is a challenging task. Nowadays, *in vitro* data[7,8] utilize idealised geometries and simplified dynamics for the study of microscopic flow because of the limitation in the fabrication of complex microvessels. It allows accurate control of the experimental parameters and a resolution which cannot be reached by *in vivo* experiments. However, because of the relatively large number of red blood cells, cell-to-cell interactions and local rheological properties were unable to identify. To study blood flow in bifurcations experimentally, people develop novel biomedical microdevices[9–11] and developed new image analysis method for analyzing microcirculatory videos[12]. A thorough review of recent development and progress on passive blood plasma separation utilizing microdevices is provided by Tripathi *et al.*[13]. The technique is based on blood properties and design of the microdevices, and shows great potential in biological as well as clinical researches.

[1]Department of Mathematics, Nanjing University of Aeronautics and Astronautics, Nanjing 210016, China. [2]School of Electronic Science and Engineering, Nanjing 210093, China. Correspondence and requests for materials should be addressed to T.W. (email: twang@nuaa.edu.cn) or Z.X. (email: zwxing@nju.edu.cn)

Despite the fact that the majority of the experimental and numerical studies have been conducted in straight blood vessels, more and more have investigated the dynamics of red blood cells in micro-bifurcations[14–20]. It has been widely accepted that the studies of bifurcating arterioles are more important because the microcirculation system consists of numerous bifurcations which form a closed network. In[21–23], experimental and/or numerical studies were carried out for aggregating and non-aggregating blood flow in a bifurcation. Because the size of the red blood cells is non-negligible comparing to the diameter of the blood vessels, the distribution of the cells is not homogeneous in arteriole scale. In refs 24–27, numerical methods such as immersed boundary-lattice Boltzmann method and dissipative particle dynamics model have been utilized to study the red blood cell partitioning and blood flux redistribution in micro-bifurcations. Wall shear stress in bifurcation was studied numerically in[28–30] because currently there are hypotheses which relate wall shear stress variations to plaques in arterial bifurcations. Due to their deformability, red blood cells migrate away from vessel walls, resulting in a region of reduced cell concentration close to the walls, namely cell free layer. Cell free layer has been measured and investigated numerically in complex geometries, such as contractions and bifurcations[18,31,32]. Interaction of red and white blood cells have also been investigated numerically[33] in bifurcating vessel networks.

Although studies have been conducted for the phenomena of blood flow in microcirculatory network for many years, it remains far from completely understood. Much less is known about the rheological behavior of red blood cells in microvessels of complex geometries, such as bifurcations. In complex vessels, interactions between red blood cells and microvessel walls strongly affect the flow behavior. Furthermore, existing simulation studies of microscopic scale on the cell distribution (or separation) in bifurcated blood vessels are inadequate, especially studies considering cell-to-cell interactions and physiologically relevant hematocrit contents. Therefore, one of the most important tasks of blood hemodynamics study is to investigate in detail and to predict blood characteristics occurring at diverging and converging bifurcations.

The present study used numerical simulations to better characterise local hemodynamics in the bifurcations with dimensions geometrically similar to arterioles. Microscopic blood flow and behavior of multiple red blood cells in curved and bifurcated vessels were simulated using two-dimensional computational models. The red blood cells were modeled as membrane particles connected by springs with stretch and compression resistance and bending rigidity to account for the stiffness of the cell membrane[4,34]. In the computation of the fluid flow, the fictitious domain method was utilized to solve the Navier-Stokes equations. The immersed boundary method was adopted to couple the cell deformation with the fluid. Motion and deformation of multiple red blood cells flowing in a microchannel with symmetric and asymmetric diverging and converging bifurcation were reported under a number of flow conditions and the influences of parameters, such as initial cell position, shape of the cells, hematocrit, and cell stiffness were analysed. Although three-dimensional simulations have been conducted in studies of bifurcated microvessels, two-dimensional simulations are still helpful as a tool to aid conceptual understanding of large systems where three-dimensional simulations are extremely costly. A systematic two-dimensional study may be used to predict the partition of hematocrit in micro-bifurcations and the effect of bifurcations on the microscopic blood flow dynamics.

## Results

We studied the hydrodynamic behavior of multiple red blood cells in a horizontal channel filled with Newtonian fluid with a bifurcation at the central part of the vessel. Two geometries of the bifurcation have been considered, namely symmetric and asymmetric. We performed a series of simulations to study red blood cell deformation, flow field, and cell-structure interactions as the red blood cells traversed the bifurcated vessel. Red blood cells modeled by the elastic spring model (see Methods) was initially placed vertically in the channel with a horizontal distance 2.5 $\mu$m between the leftmost cell center and the left inlet. The gap between the centers of the two adjacent cells was fixed at 3.75 $\mu$m (for 8 cells, 16 cells and 44 cells) or 2.5 $\mu$m (for 80 cells). The fluid flow was generated from left to right by a constant pressure drop. One stream approaching the bifurcation divided into two flows in daughter vessels of the bifurcation, and then they entered the confluence and merge together into a single flow. Afterwards, the flow was allowed to develop fully along the straight rectangular channel. At the meantime, prelocated multiple red blood cells flowed with the fluid in the channel.

**Bifurcation geometries and simulation parameters.** The microvessel in this study consisted of straight parent channel and two daughter branches representing capillaries. Figure 1(a) shows the schematic model of the bifurcation. To obtain the bifurcating geometry for the simulations, two ellipses were used to form two curved daughter branches. The bigger ellipse centers at the midpoint of the domain which was (70 $\mu$m, 20 $\mu$m) and had a major semi-axis of $a_1 = 34$ $\mu$m and a minor semi-axis $b_1 = 18$ $\mu$m. For the symmetric bifurcation geometry, the center of the smaller ellipse was located at midpoint of the domain as well, and had a major semi-axis of $a_2 = 20$ $\mu$m and a minor semi-axis $b_2 = 8$ $\mu$m. For the asymmetric bifurcation, the smaller ellipse centered at the point (70 $\mu$m, 23 $\mu$m) with the same size of axes so that a smaller branch was formed on the top and a larger branch was formed below. Using these parameters, the inlet and outlet of the daughter branches were about 12 $\mu$m and the width at the midpoint of the daughter branches was 10 $\mu$m for the symmetric bifurcation. For the asymmetric bifurcated channel, the smaller branch had a 10-$\mu$m inlet and outlet and a width of 7 $\mu$m at the midpoint, while for the larger branch, the size of the inlet and outlet was about 14 $\mu$m and the width at the midpoint was 13 $\mu$m. Initially, the red blood cells were located in the parent channel with the long axis of the cells being perpendicular to the flow direction as in Fig. 1(a).

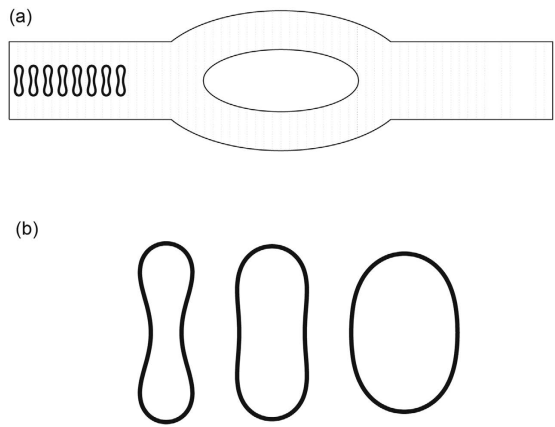

**Figure 1.** (**a**) Schematic model of a microvessel with diverging and converging bifurcations. (**b**) From left to right: red blood cell shapes obtained for reduced area $s^* = 0.481$, 0.7 and 0.9.

| Parameter | Symbol | Value |
|---|---|---|
| Blood plasma density | $\rho$ | 1.0 g/cm$^3$ |
| Blood plasma viscosity | $\mu$ | 1.2 cp |
| Radius of the circle in red blood cell model | $r_0$ | 2.8 $\mu$m |
| Number of springs in red blood cell model | $N$ | 76 |
| Membrane mass in red blood cell model | $m$ | $2.0 \times 10^{-4}$ g |
| Membrane viscosity in red blood cell model | $\gamma$ | $8.8 \times 10^{-7}$ Ns/m |
| Spring constants for red blood cell membrane | $k_l$ & $k_b$ | $3.0 \times 10^{-13}$, $1.0 \times 10^{-12}$, $3.0 \times 10^{-12}$ Nm |
| Length of the channel | $L$ | 140 $\mu$m |
| Radius of inlet (outlet) | $R$ | 20 $\mu$m |
| Grid size | $h$ | 1/64 $\mu$m |
| Time step | $\Delta t$ | $1 \times 10^{-5}$ |

**Table 1. Parameters used for the simulations.**

Three shapes of red blood cells, namely, reduced area $s^* = 0.481$, 0.7, and 0.9[4] shown in Fig. 1(b) were employed in the simulations. Among these, the shape with reduced area $s^* = 0.481$ represented health red blood cell and the shapes with $s^* = 0.7$ and $s^* = 0.9$ were for clinical abnormal cells, e.g., elliptocyte and spherocyte, respectively. In particular, the length of the cell with $s^* = 0.481$ is about 7.6 $\mu$m, which is in agreement with clinical findings.

The simulation parameters were such that the blood plasma density and plasma viscosity were fixed values. The total length of the channel was selected to be 140 $\mu$m to ensure the full development of the flow. A constant pressure gradient was imposed at the inlet and the outlet of the channel so that a fluid flow was established from left to right. We have chosen the pressure gradient based on the velocity profile for tube flow $\Delta p/L = 4 v_m \mu / R^2$ although our simulations have been done in two-dimensional channels. The maximum flow velocity $v_m$ for the tube flow was 10 cm/s (a typical value in arterioles[35]) unless otherwise stated. Other simulation parameters are listed in Table 1.

**Symmetric bifurcation.** *Effect of hematocrit.* To study the effect of hematocrit (Hct), 8 cells, 16 cells, 44 cells, and 80 cells were simulated using the model. The corresponding hematocrit values are about 3.2%, 6.4%, 17.6%, and 32% based on the two-dimensional configurations. When the hematocrit content was 3.2%, the cells formed only one file with the cell center being coaxial with the axis of the channel initially. For higher hematocrit values of 6.4%, 17.6%, and 32%, the cells formed two files which were symmetric about the horizontal centerline of the channel with the distance between the cell center and the axis of the channel being 5 $\mu$m (noncoaxial) initially. Snapshots at four instants in the simulation for the 8 cell case (Hct = 3.2%) and 16 cells (Hct = 6.4%) case are shown in Fig. 2(a–d) and Supplementary Fig. S1, respectively.

In Fig. 2(a–d), 8 red blood cells (Hct = 3.2%) formed a single file initially in the parent channel near the inlet. As time elapsed, they moved towards the outlet in the flow while deform themselves to reduce the resistance acting by the blood plasma. They slowed down and distorted the streamline when they reached the tip of the diverging bifurcation. Five cells passed through the top daughter branch while three cells passed through the lower one because of the asymmetry of the flow which is caused by the asymmetric deformation of red blood cells in the flow. It has been found that this asymmetric deformation of red blood cells in a symmetric flow occurs below a critical value of reduced area $s^* = 0.7$[36].

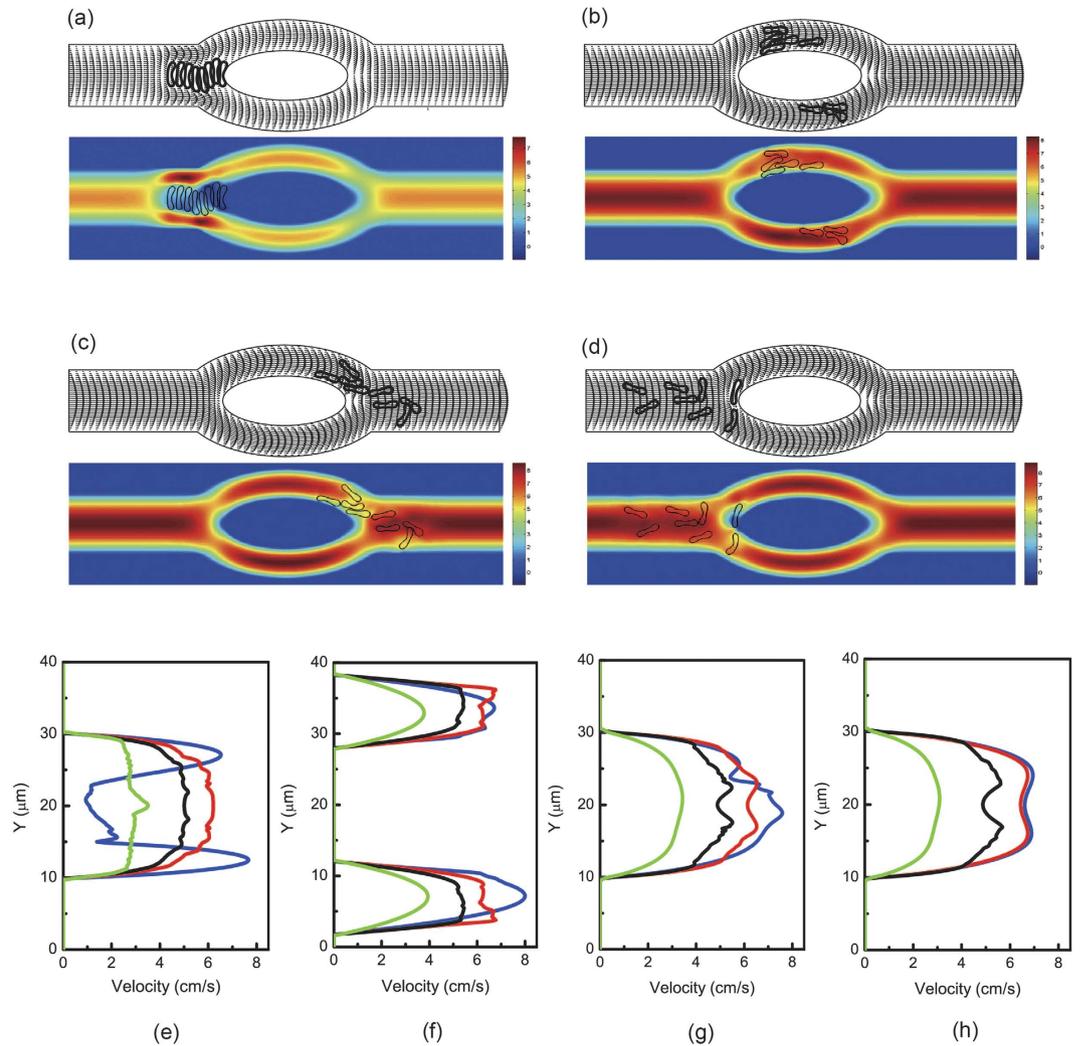

**Figure 2.** Motion of a file of 8 red blood cells (Hct = 3.2%) in the symmetric bifurcated microchannel at time instants (**a**) $t = 0.71$ ms, (**b**) $t = 1.43$ ms, (**c**) $t = 1.91$ ms, and (**d**) $t = 5.00$ ms. Velocity vectors (upper panels) and axial velocity magnitude contours (cm/s) (lower panels) are presented. The reduced area $s^* = 0.481$. The spring constant of the cell membrane was $k_l = k_b = 3.0 \times 10^{-13}$ Nm. Velocity profiles at different locations of the microvessel for different hematocrits: (**e**) at the apex of the diverging bifurcation at $t = 0.75$ ms, (**f**) at the mid cross section of the bifurcation at $t = 1.50$ ms, (**g**) at the apex of the converging bifurcation at $t = 2.00$ ms, and (**h**) at the cross section 2 $\mu$m from the right outlet at $t = 5.00$ ms. Blue line: 8 red blood cells (Hct = 3.2%); red line: 16 red blood cells (Hct = 6.4%); black line: 44 red blood cells (Hct = 17.6%); green line: 80 red blood cells (Hct = 32%).

In Supplementary Fig. S1, 16 cells (Hct = 6.4%) form two files which were symmetric about the centerline at the beginning of the simulation. Lateral migration of the red blood cells has been observed. Symmetry of the position of the cells was kept for a long time before it was totally destroyed.

The axial velocity distribution of the blood flow in the channel at various locations for the four hematocrit values is illustrated in Fig. 2(e–h). As displayed in the figure, flow propagated almost symmetrically in both daughter vessels. This symmetry was well maintained in the daughter vessels. Afterwards, fully developed blood flow was progressively obtained in the parent channel after the bifurcation. Decrease of the flow velocity with increase of hematocrit is observed in the figures.

*Effect of deformability.* Figure 3 shows the motion of 8 red blood cells (Hct = 3.2%) in the same symmetric bifurcation when the cell membrane stiffness increased by a factor of ten. The cells deformed less severely while they moved in the fluid. To further investigate the flow dynamics in the symmetric bifurcation, transit velocity of red blood cells was studied. Transit velocity $V_{tr}$ of red blood cells is defined as the average velocity of cells when they travel a same distance (length of the domain in this study) in the microvessel. The factors that affect transit velocity are the shape of the red blood cells, the cell deformability, and the friction experienced by the cell during its entry into the microchannel. For the simulation of 8 red blood cells (Hct = 3.2%), transit velocities of red blood cells in the modeled bifurcated channel were calculated and analyzed for various cases and the results are presented in Fig. 4. Each subfigure shows a scatter plot of transit velocity against membrane stiffness, which is characterized by the spring constants ($k_l$ and $k_b$) in the red blood cell model. In general, the transit velocities decrease with the increase of cell stiffness.

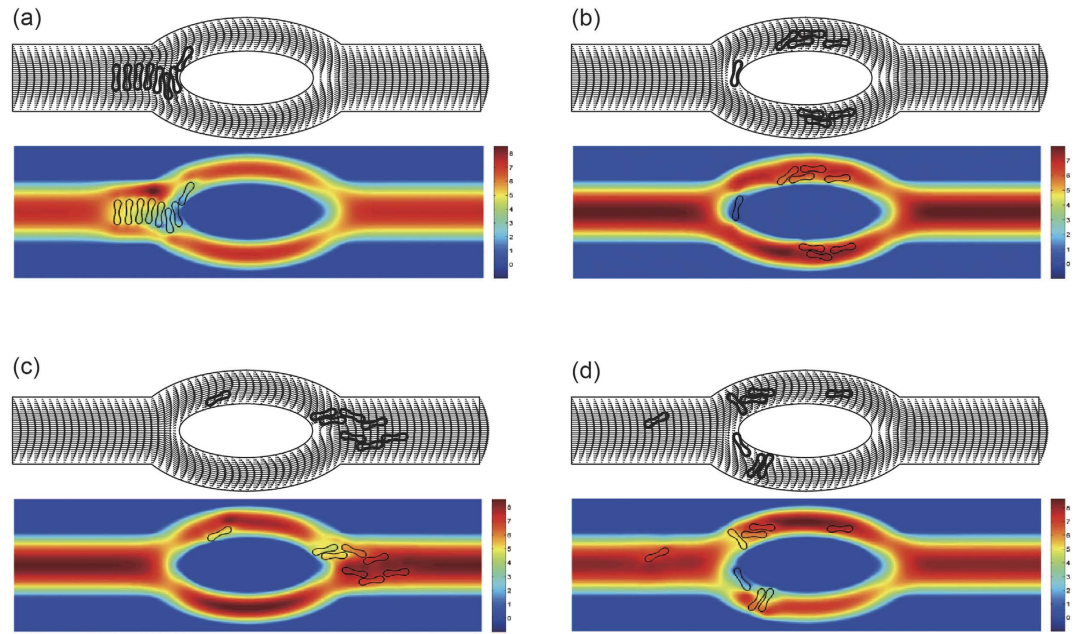

**Figure 3.** Motion of a file of 8 red blood cells (Hct = 3.2%) in the symmetric bifurcated microchannel at time instants (**a**) $t = 0.55$ ms, (**b**) $t = 1.16$ ms, (**c**) $t = 1.66$ ms, and (**d**) $t = 5.00$ ms. Velocity vectors (upper panels) and axial velocity magnitude contours (cm/s) (lower panels) are presented. The reduced area $s^* = 0.481$. The spring constant of the cell membrane was $k_l = k_b = 3.0 \times 10^{-12}$ Nm.

*Effect of initial position and shape.* The effect of initial position of the red blood cells on the transit velocity was considered and is shown in the three subfigures of Fig. 4 and they are for the cells with initial distance between cell center and the axis of the parent channel ($h_{off}$) being 0 $\mu$m (coaxial), 2.5 $\mu$m (noncoaxial), and 5.0 $\mu$m (noncoaxial), respectively. It can be seen from the graphs that the transit velocity decreased relatively slowly with respect to membrane stiffness when the cells were located off the centerline, and sharply when the cells were coaxial with the centerline of the parent channel. The reason for this is that since the pressure difference at the inlet and the outlet remained the same, when the red blood cells were coaxial, local aggregation of the cells at the apex of the diverging bifurcation increased the friction thus slowed down the transit velocity.

The effect of shape of the red blood cells on velocity field can be observed form Fig. 5. Similar flow pattern has been observed as for normal red blood cells. The red blood cells dispersed in the plasma but kept a cell free layer near the vessel wall when the simulation time was sufficiently long. The effect of red blood cell shape is also demonstrated in Fig. 4. When the cells were more swollen than their usual biconcave shape, the transit velocity decreased and this effect was more severe for the coaxial case than the other two noncoaxial cases. Moreover, the effect of initial position and shape on the cell distribution in the two daughter branches is presented in Table 2. Our results show that the red blood cell in bifurcations have similar behavior as in stenotic microvessels[5] and in bifurcations[18].

**Asymmetric bifurcation.** *Effect of hematocrit.* For the first case studied in the asymmetric bifurcation, initially 8 red blood cells (Hct = 3.2%) in a single file were placed coaxial in the parent vessel prior to the bifurcation. Blood flow in the asymmetric bifurcated channel is displayed in Supplementary Fig. S2. The results demonstrate that all red blood cells entered and passed through the branch with a larger diameter. The velocity of the flow in the larger branch is significantly higher than that in the smaller branch.

Next, simulations were carried out in the same asymmetric bifurcated channel with higher hematocrit levels, namely 6.4%, 17.6%, and 32%. The cells were in two files initially. Motions and deformations of red blood cells are presented in Supplementary Fig. S3 and Fig. 6(a–d). The red blood cells appeared to move slower in the central region of the parent channel just before the bifurcation apex. Some cells followed stream lines and ended up at the apex of the diverging bifurcation thus moved slower because of friction between the cells and the vessel wall. Local hematocrit was different for the two daughter branches. The larger branch received significantly more cells.

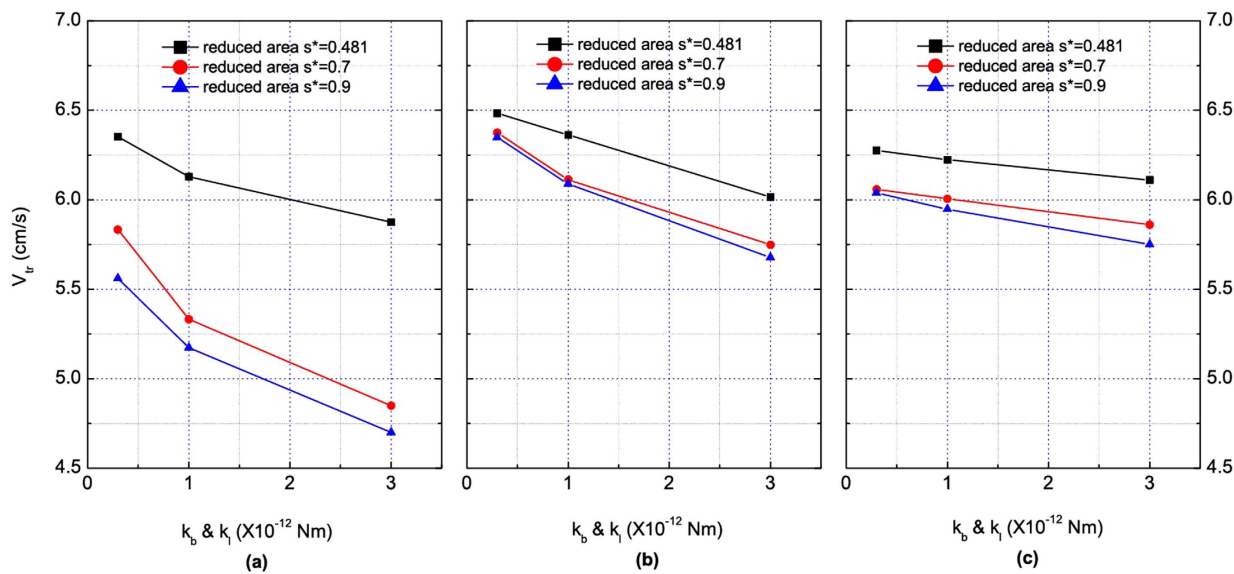

**Figure 4.** Motion of a file of 8 red blood cells (Hct = 3.2%) in the symmetric bifurcated microchannel at time instants (**a**) $t = 0.74$ ms, (**b**) $t = 1.44$ ms, (**c**) $t = 1.93$ ms, and (**d**) $t = 5.00$ ms. Velocity vectors (upper panels) and axial velocity magnitude contours (cm/s) (lower panels) are presented. The reduced area $s^* = 0.7$. The spring constant of the cell membrane was $k_l = k_b = 3.0 \times 10^{-13}$ Nm.

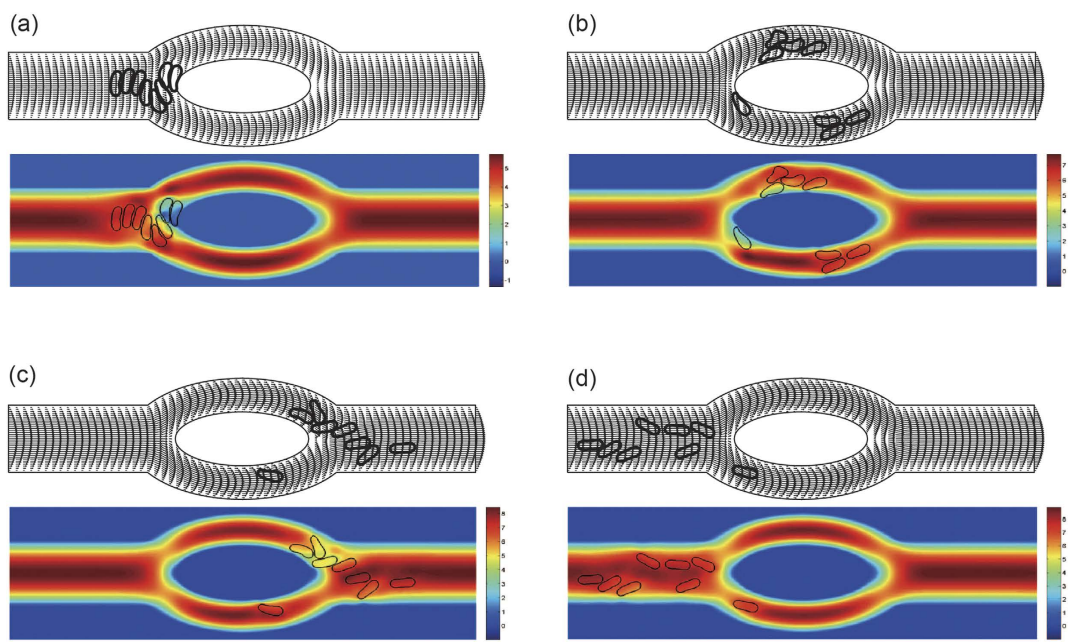

**Figure 5.** Dependence of transit velocity of erythrocytes in the symmetric bifurcated microchannel on cell membrane stiffness. (**a**) The initial distance between cell center and the axis of the channel $h_{off} = 0\,\mu$m (coaxial). (**b**) $h_{off} = 2.5\,\mu$m (noncoaxial). (**c**) $h_{off} = 5\,\mu$m (noncoaxial).

|  | $h_{off} = 0\,\mu$m | $h_{off} = 2.5\,\mu$m | $h_{off} = 5\,\mu$m |
|---|---|---|---|
| $s^* = 0.481$ | 5/3 | 8/0 | 8/0 |
| $s^* = 0.7$ | 4/4 | 8/0 | 8/0 |
| $s^* = 0.9$ | 4/4 | 8/0 | 8/0 |

**Table 2.** Effect of initial position and shape on red blood cell distributions (number of cells in the upper branch/number of cells in the lower branch) in the symmetric bifurcation.

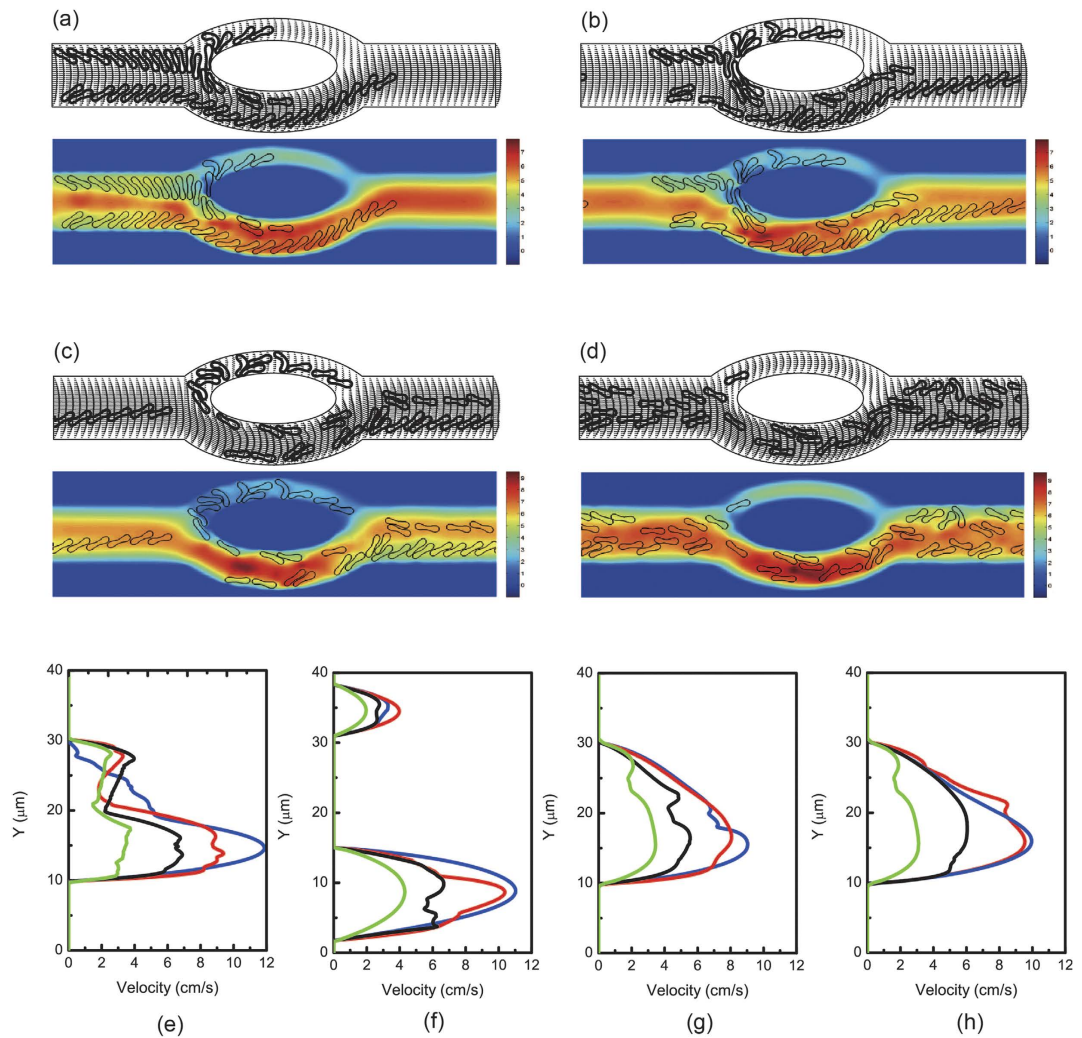

**Figure 6.** Motion of a file of 8 red blood cells (Hct = 3.2%) in the asymmetric bifurcated microchannel at time instants (**a**) $t = 0.75$ ms, (**b**) $t = 1.54$ ms, (**c**) $t = 2.40$ ms, and (**d**) $t = 5.00$ ms. Velocity vectors (upper panels) and axial velocity magnitude contours (cm/s) (lower panels) are presented. The reduced area $s^* = 0.481$. The spring constant of the cell membrane was $k_l = k_b = 3.0 \times 10^{-13}$ Nm. Velocity profiles at different locations of the microvessel for different initial cell positions: (**e**) at the apex of the diverging bifurcation at $t = 0.75$ ms, (**f**) at the mid cross section of the bifurcation at $t = 1.50$ ms, (**g**) at the apex of the converging bifurcation at $t = 2.40$ ms, and (**h**) at the cross section 2 $\mu$m from the right outlet at $t = 5.00$ ms. Blue line: $h_{off} = 0$ $\mu$m (coaxial) initially; red line: $h_{off} = 2.5$ $\mu$m (noncoaxial) initially; black line: $h_{off} = 5$ $\mu$m (noncoaxial) initially.

At the apex of the converging bifurcation, a cell free region was formed. It can also be observed that some red blood cells aggregated into clusters of two or several cells and these aggregates may dissociate in the flow.

Conversely to symmetric bifurcation, Fig. 6(e–h) shows that at the apex of the diverging bifurcation, two asymmetric peaks were produced. However, two streams entered the confluence and merged together at the apex of the converging bifurcation. In general, increase of local hematocrit lowered flow velocities and difference in flow velocities was more pronounced in the parent channel and the bigger daughter branch.

*Effect of cell deformability.* To investigate the dependence of the deformation of the cells when the red blood cells pass through the bifurcated channel on the membrane stiffness, we considered three characteristic values of the bending resistance as showed in Table 1 while other parameters were unchanged. As observed in Fig. 7(a–d), the less deformable red blood cell passed through the bifurcations with smaller deformation. Rigid cells obstructed the entrance to the low-flow daughter vessel. They blocked the vessel thus moved slower comparing to the more deformable cells.

The effect of membrane stiffness on the velocity profiles (Fig. 7(e–h)) of red blood cells was that the velocity decreased with the increase of the stiffness of the cell membrane. Effect of velocity on the membrane stiffness is more pronounced in the parent channel after the confluence than before the diverging bifurcation apex and in daughter branches. In Fig. 7(e), no major difference in the velocity profiles is observed. However, blood velocity at the mid location of the larger daughter branch has a slight decrease when the deformability of the cell membrane increases. These results agree with the experimental observations in[3].

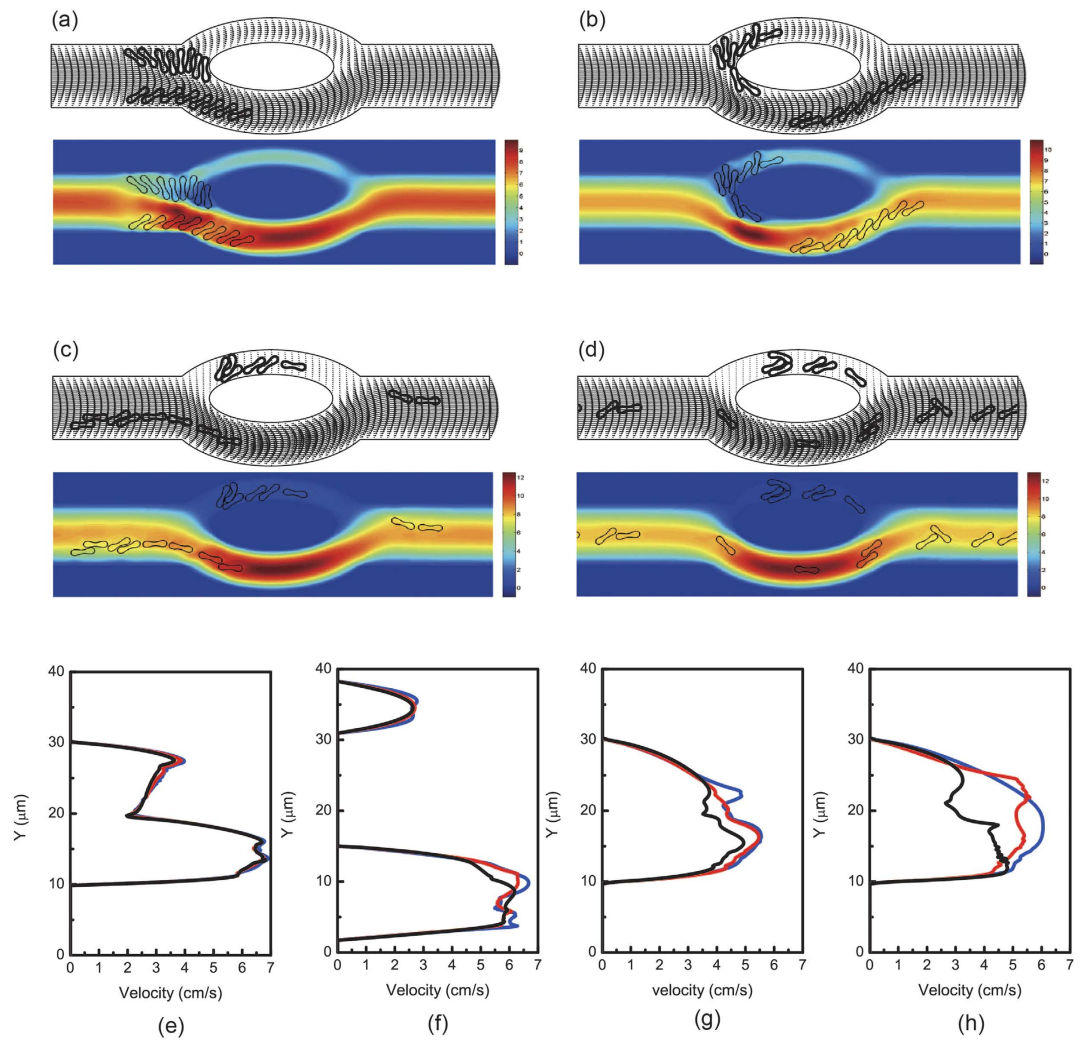

**Figure 7.** Motion of two files of 44 red blood cells (Hct = 17.6%) in the asymmetric bifurcated microchannel at time instants (**a**) $t = 1.20$ ms, (**b**) $t = 1.80$ ms, (**c**) $t = 2.40$ ms, and (**d**) $t = 5.00$ ms. Velocity vectors (upper panels) and axial velocity magnitude contours (cm/s) (lower panels) are presented. The reduced area $s^* = 0.481$. The spring constant of the cell membrane was $k_l = k_b = 3.0 \times 10^{-13}$ Nm. Velocity profiles at different locations of the microvessel for different hematocrits: (**e**) at the apex of the diverging bifurcation at $t = 1.20$ ms, (**f**) at the mid cross section of the bifurcation at $t = 1.80$ ms, (**g**) at the apex of the converging bifurcation at $t = 2.40$ ms, and (**h**) at the cross section 2 $\mu$m from the right outlet at $t = 5.00$ ms. Blue line: 8 red blood cells (Hct = 3.2%); red line: 16 red blood cells (Hct = 6.4%); black line: 44 red blood cells (Hct = 17.6%); green line: 80 red blood cells (Hct = 32%).

|  | $h_{off} = 0 \mu$m | $h_{off} = 2.5 \mu$m | $h_{off} = 5 \mu$m |
|---|---|---|---|
| $s^* = 0.481$ | 0/8 | 0/8 | 3/5 |
| $s^* = 0.7$ | 0/8 | 1/7 | 5/3 |
| $s^* = 0.9$ | 0/8 | 0/8 | 4/4 |

**Table 3. Effect of initial position and shape on red blood cell distributions (number of cells in the smaller branch/number of cells in the bigger branch) in the asymmetric bifurcation.**

*Effect of initial position and shape.* Eight normal red blood cells (Hct = 3.2%) in a single file were initially located noncoaxial in the parent channel. The cells were moved vertically up 5 $\mu$m so that the cells were more closer to the top wall. Unlike the first case, red blood cells in this configuration (Fig. 8(a–d)) were close enough to the smaller vessel to enter it. Three out of eight cells squeezed into the top smaller bifurcation with considerable deformation. The different initial position of cells indicates different initial velocities. Red blood cells located far from centerline move across streamline, i.e., they undergo lateral migration and eventually end up on the streamlines which are closer to the centerline of the channel, thus leaved a cell free layer close to the vessel walls as well as a cell free region around the tip of the confluence.

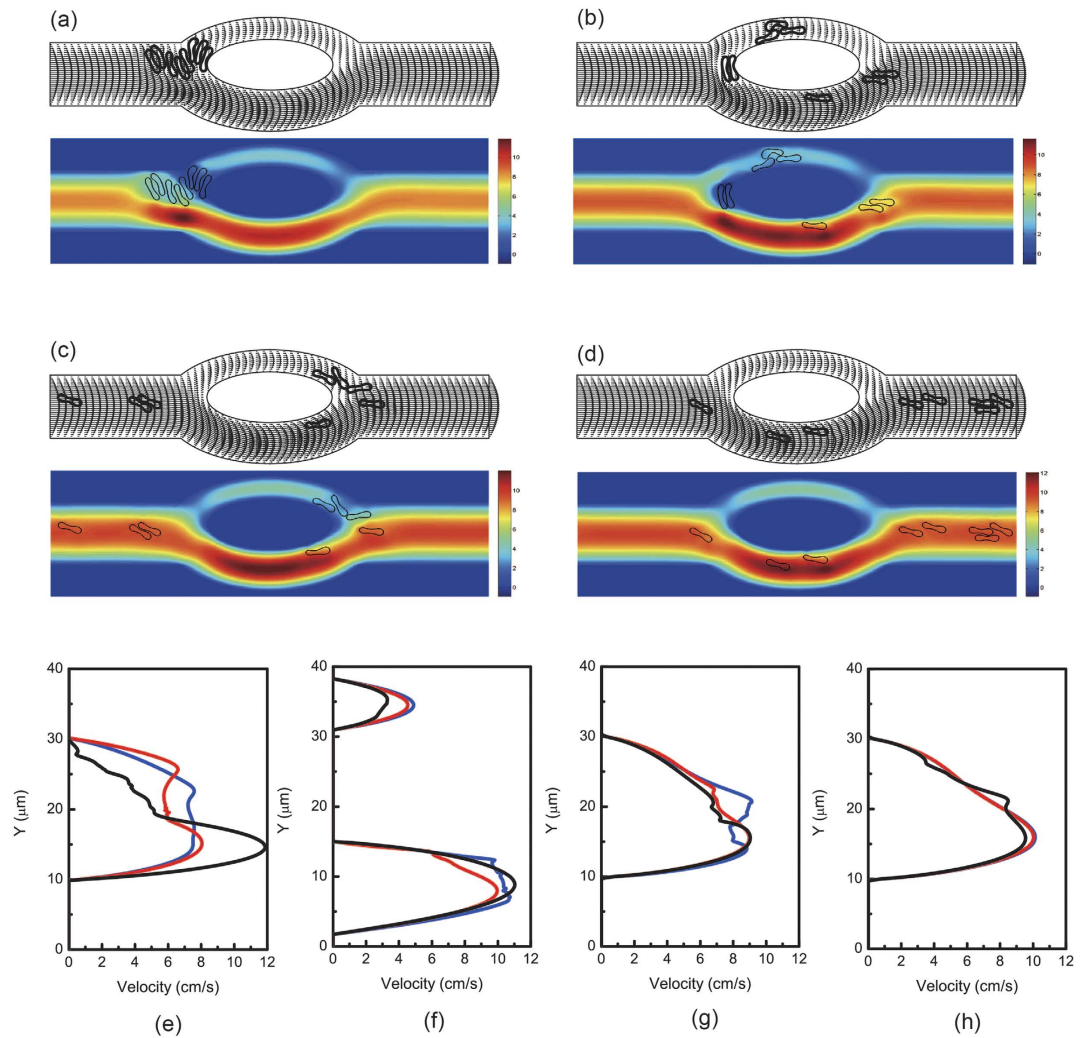

**Figure 8.** Motion of two files of 16 red blood cells (Hct = 6.4%) in the asymmetric bifurcated microchannel at time instants (**a**) $t = 0.60$ ms, (**b**) $t = 1.21$ ms, (**c**) $t = 2.40$ ms, and (**d**) $t = 5.00$ ms. Velocity vectors (upper panels) and axial velocity magnitude contours (cm/s) (lower panels) are presented. The reduced area $s^* = 0.481$. The spring constant of the cell membrane was $k_l = k_b = 3.0 \times 10^{-12}$ Nm. Velocity profiles at different locations of the microvessel for different cell stiffness: (**e**) at the apex of the diverging bifurcation at $t = 1.20$ ms, (**f**) at the mid cross section of the bifurcation at $t = 1.80$ ms, (**g**) at the apex of the converging bifurcation at $t = 2.40$ ms, and (**h**) at the cross section 2 $\mu$m from the right outlet at $t = 5.00$ ms. Blue line: $k_l = k_b = 3.0 \times 10^{-13}$ Nm; red line: $k_l = k_b = 1.0 \times 10^{-12}$ Nm; black line: $k_l = k_b = 3.0 \times 10^{-12}$ Nm.

In Fig. 8(e–h), we display the axial velocity distribution of the blood flow in the channel at various locations in the presence of red blood cells. Figure 8(e–h) shows that two streams enter the confluence and merge together after the bifurcation. Two peaks in the velocity profile can be seen. In the straight channel after the bifurcation, the flow was developed to obtain a single peak in the velocity profile.

Similar simulations have been carried out for the cells with reduced area $s^* = 0.7, 0.9$ as well. The dependence of cell distribution on the initial position and shape is presented in Table 3.

**Comparison with experiments.** Although numerical simulations may offer a cheap and robust way to investigate microscopic blood flow behavior in arterioles, comparison with experimental observations is still needed to validate the results found in the mathematical model.

Lateral migration of red blood cells in arterioles is a crucial factor that causes the formation of cell free region near the vessel wall. In our simulations, there were two major types of cell depletion regions observed, namely the cell free layer near the vascular wall in the parent channel and the cell poor region at the confluence of the daughter branches. The estimation of cell free layer thickness in the parent vessel for the four hematocrit contents is given in Fig. 9(a) and the cell free layer thickness was found to decrease with the increase of the hematocrit values.

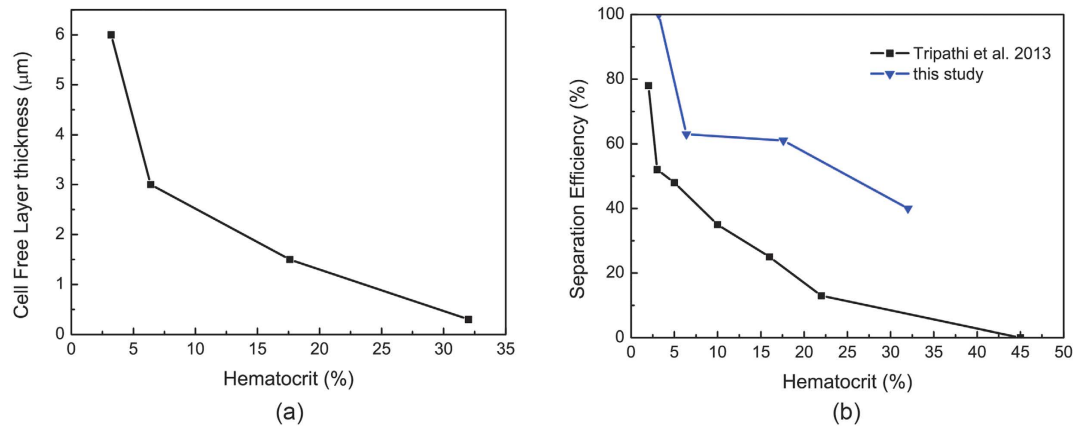

**Figure 9.** (**a**) Cell free layer thickness in the parent vessel for different hematocrit levels. (**b**) Separation efficiency of asymmetric bifurcated vessel for different hematocrit levels in comparison with experiments (T-channel, flow ratio = 8)[9].

Near the tip of the converging bifurcation, the streamlines were distorted which causes the cells moving away from the confluence and the cell free region developed here was also noted to shrink as the hematocrit increased. These results are qualitatively consistent with experimental findings[9–11].

In analogy to the definitions in the work of Prabhakar A. et al.[10], the flow ratio in this study is defined as the flow rate in the parent channel over the flow rate in the smaller branch for the asymmetric bifurcation, while the separation efficiency is defined as percentage of the number of cells passing through the bigger branch with respect to the total cells in the feed blood. The flow ratio for the asymmetric bifurcated vessel in this study was estimated as about 4:1 to 8:1 for different values of hematocrit. The separation efficiency was then compared with the experimental findings with a similar flow ratio of 8:1[10] and the results are presented in Fig. 9(b). A decrease in separation efficiency was observed with an increase in hematocrit and qualitative agreement exists between the simulations and the experiments. The discrepancy between the simulation values and experimental data may be caused by the different geometry and parameters used in the simulations and experiments. In addition, the simulations were two-dimensional while the experiments were three-dimensional studies.

## Discussion

The human circulatory system is a complicated, branched network of blood vessels of various sizes. In the present paper, we reported numerical investigation on the flow of multiple red blood cells with cell-to-cell interactions in a microchannel with symmetric and asymmetric bifurcations. The micro-scale simulations reveal a complex flow behavior of red blood cells in bifurcated microchannels. The partition of red blood cells in two asymmetric daughter branches was investigated. The dependence of motion and deformation of red blood cells on the channel geometry, hematocrit, and deformability of the cell membrane was also analysed.

In general, red blood cells migrated towards the centerline of the blood vessel downstream of the bifurcation even when the upstream position of the cells was coaxial with the microchannel. The local increase of hematocrit in the region around the apex of the diverging bifurcation caused the lower velocities in this region. Flow propagated almost symmetrically or totally asymmetrically in the two daughter vessels depended on the geometry of the bifurcation. Red blood cells hydrodynamically interacted with each other especially for the higher hematocrit cases. The distribution of red blood cells in the two daughter vessels was not uniform because the cells have a higher tendency to enter the nearest branch or the branch with the higher flow rate. It is important to note that red blood cells tend to slow down and aggregate around the apex of the diverging bifurcation, however form a cell-depleted zone immediately in the vicinity of the tip of the confluence. The geometry of the microchannel, the hematocrit percentage, as well as the deformability of the cell membrane has an influence on the formation and the size of the depleted zones.

Increase in hematocrit of the parent channel increased the local hematocrit in daughter vessels, thus increased the interaction between the red blood cells. The flow velocities of both parent channel and branches were both reduced. However, the fraction of the red blood cells entering the smaller branch for the asymmetric bifurcation was observed to increase with the increase of the source hematocrit. In general, the shape effect on the transit velocities of the red blood cells was such that more swollen cells move slower regardless the initial position of the cells in the parent channel.

Increase in cell stiffness decreased transit velocity of the red blood cells of all shapes. The effect was significant when the position of the cells was coaxial before they entered the bifurcation. When the cells were closer to the vessel wall initially, this effect became less profound. For the asymmetric bifurcation, decrease in blood flow velocity was also observed with the increase of cell membrane constants. The tendency of entering the low flow rate branch increase with the decrease in cell deformability. Blood velocity was found to depend on the deformability of the cell and decrease with the increases of the stiffness of the membrane. The flow velocity increases with the increase of the plasma viscosity.

The model results are qualitatively consistent with experimental findings on cell free region and hematocrit distributions in bifurcated vessels. The numerical simulations presented in this study yield prediction of the cell volumetric fraction in the daughter branches, which can be utilized to predict the effective blood viscosity. The results will provide much needed deeper insight into the working of the human microcirculation which will help combat disease and will aid in the development of synthetic blood substitute. In order to gain a deeper understanding of our findings, three-dimensional study will be conducted in the near future.

## Methods

We considered a two-dimensional microvessel with diverging and converging bifurcations and employed numerical simulations to study the rheological behavior of multiple red blood cells in the blood flow. In this study, blood was assumed to be a suspension of red blood cells in an incompressible, Newtonian fluid with constant density and viscosity. In order to simulate the blood flow and the fluid-cell interactions in this irregular-shaped domain, the fictitious domain method was combined with the immersed boundary method and the red blood cells have been modeled by the spring model.

### The fictitious domain formulation.

The flow region we studied was a bifurcated microchannel for which the regular structured mesh was not applicable at the boundary of the region. Thus, we adopted the fictitious domain method because in this method the irregular-shaped domain is extended to regular-shaped so that a simple structured mesh instead of unstructured mesh can be used, which substantially reduces computational complexity of the algorithm. The fictitious domain method and its applications to fluid flow problems have been extensively described[37,38]. To employ the fictitious domain method, the flow region $\Omega_f$ was embedded in a rectangular domain denoted by $\Omega$. Then the fluid flow containing red blood cells was solved in the bigger domain $\Omega$, and the no-flow condition in the solid region was treated as constraints. Therefore, the governing equations for the modeled problem were the following extended Navier-Stokes equations:

$$\rho \left[ \frac{\partial \mathbf{u}}{\partial t} + \mathbf{u} \cdot \nabla \mathbf{u} \right] = -\nabla p + \mu \Delta \mathbf{u} + \mathbf{f}, \quad \text{in } \Omega_f, \tag{1}$$

$$\nabla \cdot \mathbf{u} = 0, \quad \text{in } \Omega_f, \tag{2}$$

$$\mathbf{u} = \mathbf{0}, \quad \text{in } \Omega \backslash \Omega_f, \tag{3}$$

where $\mathbf{u}(x, t)$ and $p$ are the fluid velocity and pressure anywhere in the flow; $\rho$ is the fluid density; $\mu$ is the fluid viscosity. The boundary conditions were such that, on $\partial \Omega_f$, a no-slip condition was applied and, at the inlet and outlet of the channel, a periodic flow condition was enforced. A detailed description of the solution method of Eqs (1)–(3) can be found elsewhere[37,38].

### Immersed boundary method.

In this study, the fluid-cell interaction was dealt with by the immersed boundary method developed by Peskin[39]. Based on this method, the boundary of the deformable object is easily calculated by the following scheme: first, the force located at the immersed boundary node $\mathbf{X} = \{X_1, X_2\}$ affects the nearby fluid mesh nodes $x = \{x_1, x_2\}$ through a discrete $\delta$ function:

$$\mathbf{F}(x) = \sum_x \mathbf{F}(\mathbf{X}) D_h(\mathbf{X} - x), \quad \text{for } |\mathbf{X} - x| \leq h, \tag{4}$$

where $h$ is the uniform finite element mesh size and

$$D_h(\mathbf{X} - x) = \delta_h(X_1 - x_1) \delta_h(X_2 - x_2), \tag{5}$$

with the one-dimensional discrete $\delta$ functions being:

$$\delta_h(z) = \begin{cases} \frac{1}{4h}\left(1 + \cos\left(\frac{\pi \cdot z}{2h}\right)\right) & \text{for } |z| \leq 2h, \\ 0 & \text{for } |z| > 2h. \end{cases} \tag{6}$$

The force in Eq. (4) is part of the external force term of Eq. (1); next, the movement of the immersed boundary node is affected by all the nearby fluid mesh nodes through the same discrete $\delta$ function:

$$\mathbf{U}(\mathbf{X}) = \sum h^2 \mathbf{u}(x) \cdot D_h(\mathbf{X} - x) \text{ for } |\mathbf{X} - x| \leq 2h. \tag{7}$$

Finally, after each time step $\Delta t$, the position of the immersed boundary node is updated by

$$\mathbf{X}_{t+\Delta t} = \mathbf{X}_t + \Delta t \mathbf{U}(\mathbf{X}_t). \tag{8}$$

### Red blood cell model.

Current models describe the red blood cell as Newtonian fluid (called cytoplasm) enclosed by the cell membrane. Among these models, elastic membrane models[40,41] focus on the elastic property of the membrane and try to reproduce it through the strain energy functions. On the other hand, spring models[4,34,42–44] depict red blood cell membrane with a network of springs. In this paper, we adopted the spring model introduced in[34] and modeled individual red blood cell as cytoplasm enclosed by a membrane represented by a finite number of membrane particles connected by springs. Based on the model, elastic energy stores in the spring due to the change of the length $l$ of the spring with respected to its reference length $l_0$ and the change in angle $\theta$ between two neighboring springs. The total elastic energy of the red blood cell membrane, $E = E_l + E_b$, is the sum of the total elastic energy for stretch and compression and the total elastic energy for bending which, in particular, are:

$$E_l = \frac{k_l}{2}\sum_{i=1}^{N}\left(\frac{l_i - l_0}{l_0}\right)^2 \tag{9}$$

and

$$E_b = \frac{k_b}{2}\sum_{i=1}^{N}\tan^2(\theta_i/2). \tag{10}$$

In Eq. (9) and Eq. (10), $N$ is the total number of the spring elements, and $k_l$ and $k_b$ are spring constants for changes in length and bending angle, respectively. $\theta_i$ is the angle formed by the $i$th particle and its two adjoined springs.

The shape of the red blood cell is obtained using the elastic spring model based on the minimum-energy principle. Initially, the cell is assumed to be a circle of radius $r_0$. The circle is discretized into $N$ membrane particles so that $N$ springs are formed by connecting the neighboring particles. The shape change is stimulated by reducing the total area of the circle $s_0$ through a penalty function[34]:

$$\Gamma_s = \frac{k_s}{2}\left(\frac{s - s_e}{s_e}\right)^2, \tag{11}$$

where $s$ and $s_e$ are the time-dependent area and the equilibrium area of the red blood cell, respectively. The total elastic spring energy $E$ is modified as quasi-energy $E_q = E + \Gamma_s$ and the force acting on the $i$th membrane particle now is:

$$\mathbf{F}_i = -\frac{\partial E_q}{\partial \mathbf{r}_i}. \tag{12}$$

The force generated by the deformation of the membrane is treated as a part of the external force term of Eq. (1). When the area is reduced, each membrane particle moves according to the following equation of motion:

$$m\ddot{\mathbf{r}}_i + \gamma\dot{\mathbf{r}}_i = \mathbf{F}_i \tag{13}$$

Here, (˙) denotes the time derivative; $m$ represents the mass of the membrane; $\gamma$ is a friction for numerical calculation. The quasi energy decreases with the time elapse. The final shape of the red blood cell is obtained as the quasi energy is minimized.

By taking into consideration the nonextensible property of the membrane, the values of spring constants are set as $k_l = k_b$. The value of penalty coefficient $k_s$ is $k_b \times 10^4$. An initial circular shape is transformed into its final stable shape associated with a minimal energy for a given reduced area $s^* = s_e/s_0$ regardless of the choice of $k_b$. The biconcave shape obtained for $s^* = 0.481$ resembles the normal physiological shape of the red blood cell very well[4].

The bending constant is closely related to the rigidity of the membrane. A higher $k_b$ results in a less deformable cell. Thus deformability of normal and hardened red blood cells can be modeled by varying spring constants for these two resistances.

## Acknowledgements

Z.W. Xing thanks the support by the State Key Program for Basic Researches of China under Grant No. 2014CB921103.

## Author Contributions

T.W. designed and performed all the simulations with support from U.R. and Z.X.; T.W. and Z.X. interpreted the results and wrote the manuscript.

## Additional Information

**Supplementary information** accompanies this paper at http://www.nature.com/srep

**Competing financial interests:** The authors declare no competing financial interests.

**How to cite this article**: Wang, T. *et al.* A micro-scale simulation of red blood cell passage through symmetric and asymmetric bifurcated vessels. *Sci. Rep.* **6**, 20262; doi: 10.1038/srep20262 (2016).